\documentclass[%
 aip,
 jmp,%
 amsmath,amssymb,
 reprint,%
]{revtex4-1}

\usepackage{graphicx}
\usepackage{dcolumn}
\usepackage{bm}
\usepackage{color}
\usepackage{ulem}
\usepackage{braket}

\begin{document}


\title[Voltage Tunable Plasmon Propagation in Dual Gated Bilayer Graphene]{Voltage Tunable Plasmon Propagation in Dual Gated Bilayer Graphene}

\author{Seyed M. Farzaneh}
 \email{farzaneh@nyu.edu}
 \affiliation{ 
Department of Electrical and Computer Engineering, New York University, Brooklyn, NY 11201, USA}
\author{Shaloo Rakheja}
 \email{shaloo.rakheja@nyu.edu}
\affiliation{ 
Department of Electrical and Computer Engineering, New York University, Brooklyn, NY 11201, USA}

\date{\today}

\begin{abstract}
In this paper, we theoretically investigate plasmon propagation characteristics in AB and AA stacked bilayer graphene (BLG) in the presence of energy asymmetry due to an electrostatic field oriented perpendicularly to the plane of the graphene sheet.
We first derive the optical conductivity of BLG using the Kubo formalism incorporating energy asymmetry and finite electron scattering. All results are obtained for room temperature (300K) operation.
By solving Maxwell's equations in a dual gate device setup, we obtain the wavevector of propagating plasmon modes in the transverse electric (TE) and transverse magnetic (TM) directions at terahertz frequencies.
The plasmon wavevector allows us to compare the compression factor, propagation length, and the mode confinement of TE and TM plasmon modes in bilayer and monolayer graphene sheets and also study the impact of material parameters on plasmon characteristics.
Our results show that the energy asymmetry can be harnessed to increase the propagation length of TM plasmons in BLG.
AA stacked BLG shows a larger increase in propagation length than AB stacked BLG; conversely, it is very insensitive to the Fermi level variations. 
Additionally, the dual gate structure allows independent modulation of the energy asymmetry and the Fermi level in BLG, which is advantageous for reconfiguring plasmon characteristics post device fabrication. 
\end{abstract}

\maketitle

\section{\label{sec:introduction}introduction}
The two-dimensional (2D) material graphene offers strong light-matter interaction over a broad electromagnetic spectrum, ranging from several hundreds of GHz to optical frequencies \cite{grigorenko2012graphene}. 
An interesting and useful attribute of graphene is its ability to support the propagation of surface plasmon polaritons (SPPs).
SPPs (referred to as plasmons, hereafter)
are collective oscillations of electrons, typically occurring at the interface between a metal and a dielectric, 
that can be coupled with electromagnetic waves and enable the manipulation of light below its diffraction limit\cite{jablan2013plasmons}. 
Plasmon oscillations can occur in various metallic and semiconducting material systems including 2D electron gases \cite{allen1977observation, liu2008plasmon, theis1980plasmons}. 
Typically, plasmons have a smaller wavelength compared to the wavelength of light at the same frequency allowing significant mode compression.

Plasmons supported on a graphene sheet have a tighter confinement and a higher propagation length relative to those of surface plasmons in metals \cite{koppens2011graphene, zhou2012atomically}. Moreover, graphene can also support the propagation of transverse electric (TE) modes \cite{mikhailov2007new}, which are absent in the case of a metal/dielectric system. 
The characteristics of graphene plasmons can also be tuned via electrostatic gating or chemical doping over a wide frequency range. In particular, the electrostatic tuning of plasmons enables reconfigurable plasmonic devices with graphene -- a feature that is unique to graphene and not afforded by metal plasmonics. 

Recently, plasmon oscillations in monolayer and multilayer graphene heterostructures were experimentally imaged \cite{fei2015tunneling, menabde2016direct, chen2012optical, woessner2015highly, ju2011graphene}. These experiments pave the way to realize tunable plasmonic devices that are currently being researched and developed including waveguides\cite{vakil2011transformation}, nanoantennas\cite{jornet2013graphene}, photodetectors\cite{vicarelli2012graphene}, ocsillators\cite{rana2008graphene}, and modulators\cite{liu2011graphene}. 
From a theoretical standpoint, plasmons have been studied extensively in gated \cite{rakheja2016gate} and ungated \cite{jablan2009plasmonics, hanson2008dyadic} monolayer graphene (MLG) and also in ungated \cite{jablan2011transverse} bilayer graphene (BLG). However, not much attention has been given to plasmons in BLG structures in which the bandstructure and, therefore, optical properties can be readily tuned via electrostatic gating. 



In this paper, we develop theoretical models to study the propagation of TE and TM (transverse magnetic) plasmons in gated BLG structures for 
various material parameters of graphene and the dielectrics surrounding the graphene sheet.
We also highlight the distinguishing features of plasmon propagation in BLG and MLG.
While both MLG and BLG exhibit tunable optical properties, the tunability is significantly enhanced in BLG.
This enhancement is a result of the electric-field-induced energy asymmetry in BLG structures. 
Essentially, a vertical electric field in a metal-oxide-graphene geometry, creates an energy asymmetry between the two parallel layers of BLG \cite{castro2008bilayer, mccann2006asymmetry, mccann2007low}.
The energy asymmetry changes the bandstructure of BLG, which in turn alters its optical properties significantly \cite{ohta2006controlling, zhang2009direct}.
In addition to the energy asymmetry, we also consider other phenomena, such as the electron scattering rate and the stacking forms, that affect the optical properties of BLG.

The remainder of this paper is organized as follows. In Section \ref{sec:plasmon}, we derive the plasmon wavevector for both TE and TM polarized modes in a graphene sheet using the solutions of Maxwell's equations. We use a device setup in which the graphene sheet is modeled as an impedance boundary condition with a finite dynamical surface conductivity. 
To obtain the dynamical conductivity, we first derive the bandstructure of BLG using tight binding model in Section \ref{sec:model}. The bandstructure is obtained in terms of the energy asymmetry between the layers and for two common stacking orders, that are AB and AA stacked BLG. 
The bandstructure is used to solve for the dynamical conductivity of BLG within the Kubo formalism.
The effects of gate voltage and electrostatic screening on the carrier concentration and energy asymmetry are included using a self-consistent Hartree approximation. 
We present the results of our analysis in Section \ref{sec:results} in which we quantify the impact of physical parameters of BLG on the plasmon characteristics such as compression factor, propagation length, and mode confinement.
The key findings of the paper are summarized in Section \ref{sec:conclusion}.

\section{\label{sec:plasmon} Plasmon Waves in Graphene}
\begin{figure} [ht]
   \begin{center}
   \begin{tabular}{c} 
   \includegraphics[height=3cm]{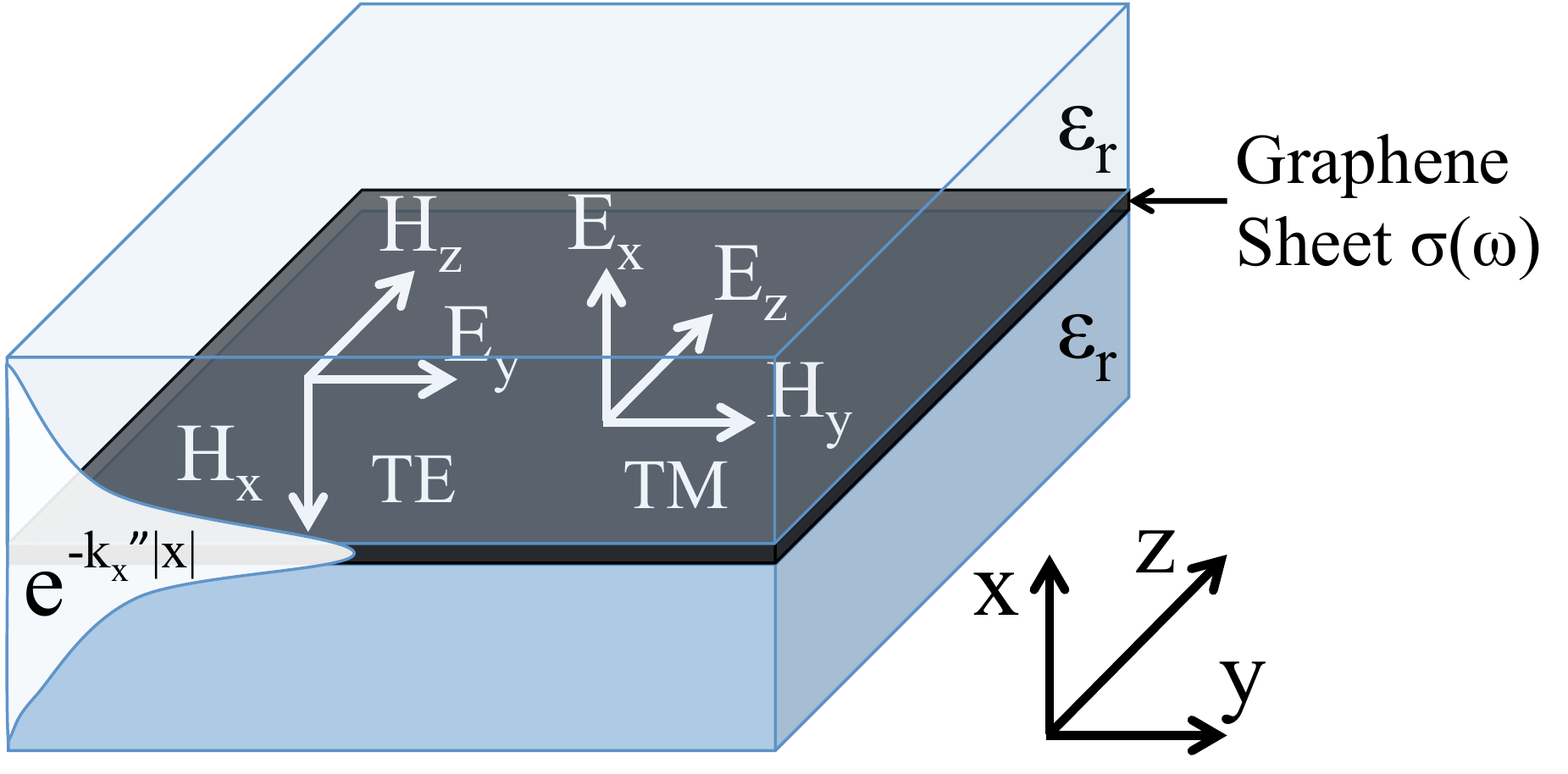} 
   \end{tabular}
   \end{center}
   \caption[example] 
   { \label{fig:plasmon-geo} 
Device setup to study the propagation of plasmons in a graphene sheet with a surface conductivity of $\sigma(\omega)$ immersed in a uniform dielectric medium ($\epsilon_r$). The components of electric and magnetic fields for TE and TM plsamon modes are labeled in the figure. Plasmons decay by $e^{-k''_x|x|}$ in the direction perpendicular to the graphene sheet.}
   \end{figure}

To obtain plasmon wavevector in graphene heterostructures, we consider the device setup shown in Fig.~\ref{fig:plasmon-geo} in which an infinite 2D sheet of graphene is immersed in a uniform, linear dielectric medium with a relative dielectric constant  of $\varepsilon_r$. 
We note that graphene in this device setup could be either a monolayer or a bilayer sheet.
The propagating electric and magnetic fields in the $z$-direction for TE-polarized plasmon mode are
\begin{equation}
\label{eq:plasmon-field-te}
\begin{split}
&\begin{cases}
{\bm E} = (E_{y_+}\hat{y})e^{-j(k_zz + k_xx)}\\
{\bm H} = (H_{x_+}\hat{x} + H_{z_+}\hat{z})e^{-j(k_zz + k_xx)}
\end{cases}\quad x>0, \\
&\begin{cases}
{\bm E} = (E_{y_-}\hat{y})e^{-j(k_zz - k_xx)}\\
{\bm H} = (H_{x_-}\hat{x} + H_{z_-}\hat{z})e^{-j(k_zz - k_xx)}
\end{cases}\quad x<0, \\
\end{split}
\end{equation}
and for TM-polarized plasmon mode are
\begin{equation}
\label{eq:plasmon-field-tm}
\begin{split}
&\begin{cases}
{\bm H} = (H_{y_+}\hat{y})e^{-j(k_zz + k_xx)},\\
{\bm E} = (E_{x_+}\hat{x} + E_{z_+}\hat{z})e^{-j(k_zz + k_xx)}\\
\end{cases}\quad x>0, \\
&\begin{cases}
{\bm H} = (H_{y_-}\hat{y})e^{-j(k_zz - k_xx)},\\
{\bm E} = (E_{x_-}\hat{x} + E_{z_-}\hat{z})e^{-j(k_zz - k_xx)}\\
\end{cases}\quad x<0, \\
\end{split}
\end{equation}
where time dependence term $e^{j\omega t}$ is dropped and $k_z (k_x)$ is the wavevector in $z$-($x$-) direction as illustrated in Fig.~\ref{fig:plasmon-geo}. 
If we plug Eqs. (\ref{eq:plasmon-field-te}) and (\ref{eq:plasmon-field-tm}) into Helmholtz equation ($\nabla^2{\bm E} + k^2_\text{light}{\bm E} = 0$), we get $k_x^2 + k_z^2 = k^2_\text{light}$, where $k_\text{light}=\omega\sqrt{\varepsilon_r\varepsilon_0\mu_0}$ is the wavevector of light in the medium. Here, $\omega$ is the radial frequency, and $\mu_0$ and $\varepsilon_0$ are the permeability and permittivity, respectively, of free space.

By solving Maxwell's equations in the device setup of Fig.~\ref{fig:plasmon-geo} and applying appropriate boundary conditions (impedance boundary with a finite surface conductivity for graphene), we obtain the following dispersion relations \cite{mikhailov2007new, hanson2008dyadic}
\begin{subequations}
\label{eq:plasmon-dispersion}
\begin{align}
\quad 1 + \frac{j\mu_0\omega\sigma(\omega)}{2\sqrt{k_z^2 - \varepsilon_r\omega^2/c^2}}=0\, , \quad \text{TE Mode}, \\
\quad 1 - \frac{j\sigma(\omega)\sqrt{k_z^2 - \varepsilon_r\omega^2/c^2}}{2\varepsilon_r\varepsilon_0\omega}=0\, , \quad \text{TM Mode},
\end{align}
\end{subequations}
where $\sigma(\omega)$ is a complex quantity that represents surface conductivity of the graphene sheet, and $c=1/\sqrt{\epsilon_0\mu_0}$ is the velocity of light in free space. 
By replacing $\sigma(\omega) = \sigma'(\omega) - j\sigma''(\omega)$ and $\eta=\sqrt{\mu_0/\varepsilon_r\varepsilon_0}$ in Eq.~(\ref{eq:plasmon-dispersion}) and rearranging the equations, we obtain the plasmon wavevectors
\begin{equation}
\label{eq:dispersion-te-detailed}
k_z = k_\text{light}\bigg(1 + \frac{\eta^2}{4} (\sigma''(\omega)^2 - \sigma'(\omega)^2) + j \frac{\eta^2}{2}\sigma'(\omega)\sigma''(\omega) \bigg)^{1/2}
\end{equation}
for TE mode and 
\begin{equation}
\label{eq:dispersion-tm-detailed}
k_z =  k_\text{light}\bigg(1 + \frac{4}{\eta^2}\frac{\sigma''(\omega)^2 - \sigma'(\omega)^2}{|\sigma(\omega)|^4} - j\frac{8}{\eta^2}\frac{\sigma'(\omega)\sigma''(\omega)}{|\sigma(\omega)|^4}\bigg)^{1/2}
\end{equation}
for TM mode. From Eqs. (\ref{eq:dispersion-te-detailed}) and (\ref{eq:dispersion-tm-detailed}), it is evident that the propagation of TE (TM) mode requires $\sigma''(\omega)$ to be negative (positive). 
This is due to the fact that for propagating plasmon modes, the imaginary part of $k_z$, which represents the propagation loss, must be positive. 

While plasmons propagate in both z-direction (longitudinal) and x-direction (transverse), they are expected to decay more rapidly in the x-direction. 
The decay of plasmons can be represented using a length scale called the propagation length. The propagation length is the distance at which the intensity of the wave degrades to $1/e$ of its initial value, where $e = 2.718$ is the base of natural logarithm.
For given wavevectors $k_z=k'_z - jk''_z$ and $k_x=k'_x - jk''_x$, the propagation length is obtained as $1/2k''_z$ and $1/2k''_x$ for longitudinal and transverse directions, respectively.
The measure of confinement of the plasmon wave to the graphene sheet can be represented by its propagation length in the transverse direction. A lower transverse propagation length means higher mode confinement on the graphene surface.  Another important characteristic of a plasmon wave is its compression factor which determines how compressed the plasmon wavelength is relative to the wavelength of light calculated in the absence of the graphene sheet. The compression factor is mathematically given as the ratio of the longitudinal phase constant to the phase constant of light $k'_z/k_\text{light}$. 

An important ingredient in obtaining the plasmon dispersion is the dynamical conductivity, $\sigma(\omega)$, which depends on the bandstructure of graphene, electron scattering rate, Fermi level, and the operating frequency. In the next section, we present the model for $\sigma(\omega)$ for AB and AA stacked BLG. For brevity, we do not include the model for $\sigma(\omega)$ in MLG. Interested readers are referred to Refs. \onlinecite{falkovsky2007space, falkovsky2007optical, falkovsky2008optical, hanson2008dyadic, horng2011drude} for detailed discussion pertaining to MLG conductivity.

\section{\label{sec:model} Model of Bilayer Graphene}
\subsection{\label{sec:band} Bandstructure and Stacking forms}
\begin{figure} [ht]
   \begin{center}
   \setlength\tabcolsep{0pt} 
   \includegraphics[width=3.4in]{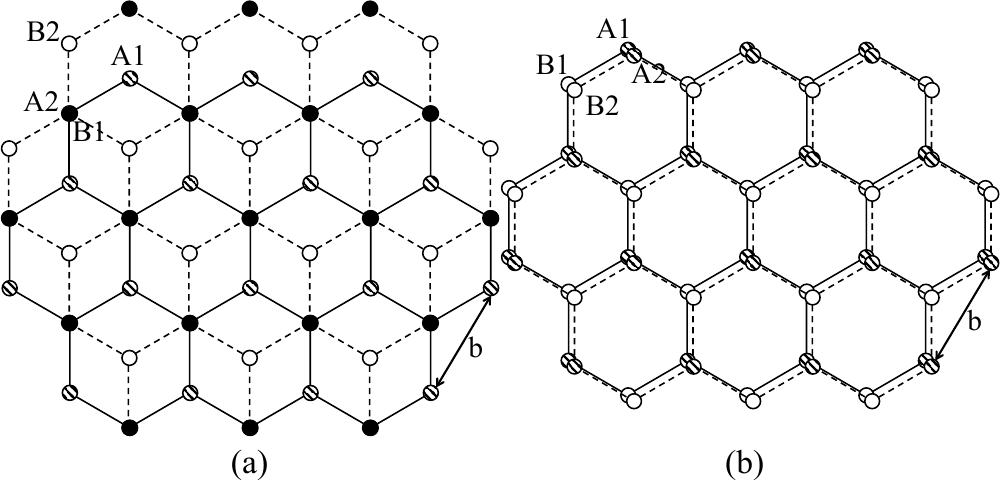}
   \end{center}
   \caption[example] 
   {Honeycomb structure of (a) AB stacked BLG and (b) AA stacked BLG. Top layer (dashed) and bottom layer (solid) are slightly displaced in (b) for clarity. The graphene lattice constant is denoted as $b=2.46\AA$.}
   \label{fig:stacking}
\end{figure}

To study the optical properties of graphene, in particular its dynamical conductivity, we start with the calculation of bandstructure of BLG.
Here, we investigate two of the most common stacking forms namely \textit{AB} (\textit{Bernal} stacked) and \textit{AA} stacked BLG. The natural stacking form of BLG is AB, which is also found in graphite. Although AA stacked BLG is likely to be metastable \cite{rozhkov2016electronic}, it is still interesting to study it from a theoretical point of view. Figure \ref{fig:stacking} illustrates the configuration of both the stacking forms. Each graphene layer consists of a triangular Bravais lattice with a basis of two atoms which we denote as `A' and `B'. In an AB staked BLG, `A' atoms of one layer align with `B' atoms of the other layer and all other atoms lie at the center of the others layer's hexagons. In an AA stacked BLG, all the atoms in one layer align exactly with other layer's atoms correspondingly. 

By utilizing tight binding model, we can write the Hamiltonian of an AB stacked BLG that is exposed to a vertical electrostatic field. The electric field introduces an energy asymmetry between the two layers, which is denoted as $\Delta$. It lowers the energy of the first layer by $\Delta/2$ and raises the energy of the second layer by the same amount. 
The Hamiltonian for an AB stacked BLG is written in the following form \cite{nicol2008optical}
\begin{equation}
\label{eq:band-ab-hamil}
\hat{H} = 
\begin{bmatrix}
-\Delta/2 & 0 &  0 & \phi^*({\bm k}) \\
0 & \Delta/2 & \phi({\bm k}) & 0 \\
0 & \phi^*({\bm k}) & \Delta/2 & \gamma_1 \\
\phi({\bm k}) & 0 & \gamma_1 & -\Delta/2
\end{bmatrix}\cdot 
\end{equation}
 \begin{figure} [ht]
   \begin{center}
   \setlength\tabcolsep{0pt} 
   \includegraphics[width=3.4in]{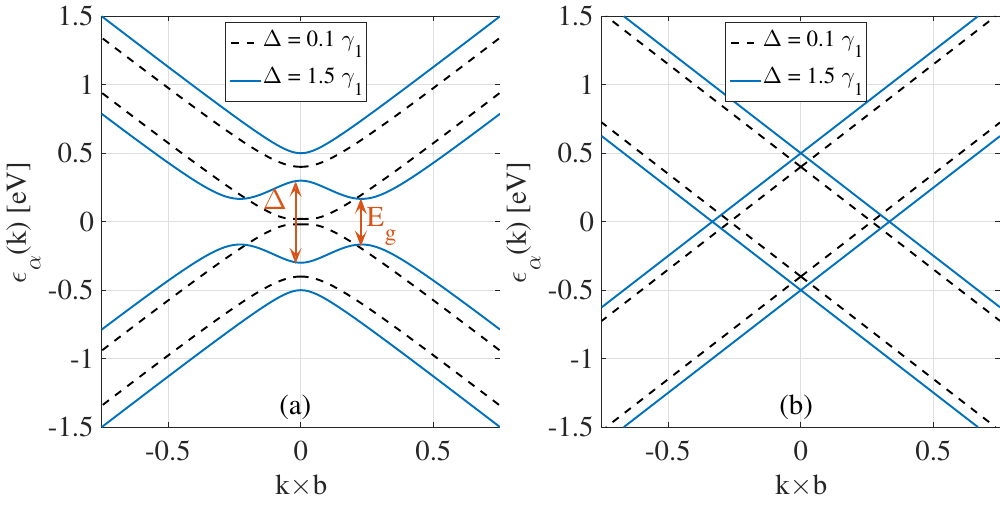}
   \end{center}
   \caption[example] 
   {Bandstructure of (a) AB stacked BLG and (b) AA stacked BLG. The bandstructure is plotted for two different values of energy asymmetry $\Delta$. In this figure $\gamma_1$ represents the interlayer hopping parameter. When both $\Delta$ and $\gamma_1$ are equal to zero, the BLG bandstructures of Eqs. (\ref{eq:band-ab-ek}) and (\ref{eq:band-aa-ek}) reduce to the energy dispersion of MLG. The energy bandgap $E_g$ occurs at $k\not=0$.}
   \label{fig:band} 
   \end{figure} 
Here, $\gamma_1=0.4$ eV is the interlayer hopping energy between $B_1$ and $A_2$ sites\cite{zhang2008determination}, and $|\phi({\bm k})|=|\hbar v_Fke^{i\theta}|$ is the energy dispersion of MLG in the vicinity of the {\textbf{K}} point of the Brillouin Zone within a radius of $3$ eV. Also $v_F=\sqrt{3}\gamma_0b/2\hbar$ denotes the Fermi velocity, $\gamma_0=3$ eV is nearest neighbor hopping energy within each layer, and $b=0.246$ nm is graphene's lattice constant.
The eigenvalues of the Hamiltonian describe the bandstructure, which in the case of AB stacked BLG is given as
\begin{equation}
\label{eq:band-ab-ek}
\begin{split}
&\epsilon_{\alpha}({\bm k}) = \\
&\pm \sqrt{\frac{\gamma_1^2}{2} + \frac{\Delta^2}{4} + |\phi({\bm k})|^2 + (-1)^{\alpha}\sqrt{\frac{\gamma_1^4}{4} + |\phi({\bm k})|^2(\gamma_1^2 + \Delta^2)}}\cdot
\end{split}
\end{equation}
Here, $\alpha=1,2$ is the subband index. We can see that for $\gamma_1=0$ and $\Delta=0$ this energy dispersion reduces to that of MLG. Figure \ref{fig:band}a shows the bandstructure of AB stacked BLG for two different values of $\Delta$. 
Contrary to MLG that exhibits a gapless linear energy dispersion relationship, AB stacked BLG shows a nonlinear energy dispersion with a finite energy bandgap.
We note that the energy bandgap ($E_g=\Delta\gamma_1/\sqrt{\Delta^2 + \gamma_1^2}$) occurs at $k\not=0$ and is smaller than $\Delta$.
Similarly, we can write the Hamiltonian of AA stacked BLG for a finite electrostatic bias as follows\cite{tabert2012dynamical}
\begin{equation}
\label{eq:band-aa-hamil}
\hat{H} = 
\begin{bmatrix}
-\Delta/2 & 0 &  \gamma_1 & \phi({\bm k}) \\
0 & \Delta/2 & \phi^*({\bm k}) & \gamma_1 \\
\gamma_1 & \phi({\bm k}) & \Delta/2 & 0 \\
\phi^*({\bm k}) & \gamma_1 & 0 & -\Delta/2
\end{bmatrix}\cdot 
\end{equation}
The eigenvalues of this matrix describe the bandstructure of AA stacked BLG, which is given as
\begin{equation}
\label{eq:band-aa-ek}
\epsilon_{\alpha}({\bm k}) = \pm \bigg(\sqrt{\gamma_1^2 + \frac{\Delta^2}{4}} +(-1)^{\alpha} |\phi({\bm k})|\bigg)\cdot
\end{equation}
The above expression shows that the energy asymmetry does not open a bandgap in the case of AA stacked BLG, and $\Delta$ can be treated as a correcting factor to $\gamma_1 \rightarrow \sqrt{\gamma_1^2 + \Delta^2/4}$. Figure \ref{fig:band}b shows the bandstructure of AA stacked BLG for two different values of $\Delta$.

\subsection{\label{sec:oc} Optical Conductivity}
We can write the real part of frequency dependent conductivity $\sigma'(\omega)$ using Kubo formalism. Following procedure described in Ref. \onlinecite{nicol2008optical}, for an AB stacked BLG we have
\begin{equation}
\label{eq:oc-ab-sigma}
\begin{split}
\sigma'_{\text{AB}}(\omega) = 
&\frac{2e^2}{\omega}\int_{-\infty}^{\infty}\frac{dE}{2\pi}[f(E) - f(E + \hbar\omega)]\\
&\int\frac{d^2{\bm k}}{(2\pi)^2}|v_{\bm k}|^2 \bigg[A_{11}(E, \Delta)A_{44}(E + \hbar\omega, \Delta)\\
&+ A_{44}(E, \Delta)A_{11}(E + \hbar\omega, \Delta)\\
&+ A_{11}(E, -\Delta)A_{44}(E + \hbar\omega, -\Delta)\\
&+ A_{44}(E, -\Delta)A_{11}(E + \hbar\omega, -\Delta) \\
&+ 2(A_{13}(E, \Delta)A_{13}^*(E + \hbar\omega, -\Delta)\\
&+ A_{13}^*(E, -\Delta)A_{13}(E + \hbar\omega, \Delta)) \bigg]\cdot
\end{split}
\end{equation}
Here, $\mu$ is the Fermi level, $f(E)=1/(1 + \exp((E - \mu)/k_BT))$ is the Fermi-Dirac distribution function with $k_B$ as the Boltzmann's constant and $T$ as the temperature, and $|v_{\bm k}|$ is the velocity of electrons and is approximated by $v_F$. The inner integral is being evaluated over the {\textbf{K}} point in Brillouin zone. The $A_{ij}$ functions are the spectral representations of Green's functions. They are related to the Green's functions through following equation:
\begin{equation}
\label{eq:oc-ab-green}
G_{ij}(z) = \int_{-\infty}^{\infty}\frac{dw}{2\pi}\frac{A_{ij}(w)}{z - w}\, ,
\end{equation}
and can be written in the form of 
\begin{equation}
\label{eq:oc-ab-spectral}
A_{ij}(w, \Delta) = \sum_{\alpha=1,2} [a_{ij}(\alpha, \Delta)\delta(w-\epsilon_\alpha) + a_{ij}(\alpha, -\Delta)\delta(w+\epsilon_\alpha)]\, ,
\end{equation} 
in which $\alpha$ denotes the $\alpha^\text{th}$ subband, and $\epsilon_\alpha$ is the subband energy. The coefficients $a_{ij}(\alpha, \Delta)$ are derived from Eqs.~(\ref{eq:oc-ab-green}), (\ref{eq:oc-ab-spectral}), and the Hamiltonian through $\hat{G}^{-1} = z\hat{I} - \hat{H}$.
\begin{equation}
\label{eq:oc-ab-coeff}
\begin{split}
&a_{11}(\alpha, \Delta) = \frac{(-1)^{\alpha+1}\pi}{\epsilon_2^2 - \epsilon_1^2}\bigg( [(\frac{\Delta}{2})^2 + \gamma_1^2 - \epsilon_\alpha^2](1 - \frac{\Delta}{2\epsilon_\alpha}) \\
&+ |\phi({\bm k})|^2 (1+\frac{\Delta}{2\epsilon_\alpha})\bigg), \\
&a_{44}(\alpha, \Delta) = \frac{(-1)^{\alpha+1}\pi}{\epsilon_2^2 - \epsilon_1^2}\bigg( [(\frac{\Delta}{2})^2  - \epsilon_\alpha^2](1 - \frac{\Delta}{2\epsilon_\alpha})\\
&+ |\phi({\bm k})|^2 (1+\frac{\Delta}{2\epsilon_\alpha})\bigg), \\
&a_{13}(\alpha, \Delta) = \frac{(-1)^{\alpha}\pi}{\epsilon_2^2 - \epsilon_1^2}(1 - \frac{\Delta}{2\epsilon_\alpha})\gamma_1\phi({\bm k})\cdot
\end{split}
\end{equation}
In a similar fashion, we can write the conductivity of AA stacked BLG which is described in Ref. \onlinecite{tabert2012dynamical}. The real part of conductivity is written as follows
\begin{equation}
\label{eq:oc-aa-sigma}
\begin{split}
\sigma'_{\text{AA}}(\omega) =& \frac{2e^2}{\omega}\int_{-\infty}^{\infty}\frac{dE}{2\pi}[f(E) - f(E + \hbar\omega)]\\
&\int\frac{d^2{\bm k}}{(2\pi)^2}4|v_{\bm k}|^2 \bigg[A_{11}(E+\hbar\omega)A_{11}(E)\\  
&+ A_{13}(E+\hbar\omega)A_{13}(E)\bigg], 
\end{split}
\end{equation}
in which the spectral functions are
\begin{subequations}
\label{eq:oc-aa-spectral}
\begin{align}
A_{11}(w, \Delta) = \sum_{\alpha=1,2} \frac{\pi}{2}[\delta(w - \epsilon_\alpha) + \delta(w + \epsilon_\alpha)] \, , \\
A_{13}(w, \Delta) = \sum_{\alpha=1,2} \frac{\pi}{2}[\delta(w - \epsilon_\alpha) - \delta(w + \epsilon_\alpha)] \cdot
\end{align}
\end{subequations}

To incorporate the effect of electron scatterings in graphene, the delta functions in Eqs. (\ref{eq:oc-ab-spectral}) and (\ref{eq:oc-aa-spectral}) are broadened using Lorentzians defined as
\begin{equation}
\label{eq:oc-lorentzian}
\delta(w) = \frac{1}{\pi}\frac{\Gamma}{w^2 + \Gamma^2}\cdot
\end{equation} 
The electron scattering rate is represented by $\Gamma$ which is generally a function of energy and the bandstructure. 
In the present work, we use a constant value that represents the energy-dependent scattering rate averaged over energy.
The impact of bandstructure and energy asymmetry on the scattering rate is discussed in Appendix \ref{app:scattering}.  
For all simulations in this work, $\Gamma$ is chosen as $20$ meV, unless otherwise stated, which is a typical experimental value of scattering rate in a graphene sheet\cite{monteverde2010transport}.
The imaginary part of conductivity is derived from its real part using \textit{Kramers-Kronig} relation 
\begin{equation}
\label{eq:oc-imag}
\sigma''(\omega) = -\frac{2\omega}{\pi} \mathcal{P} \int_{0}^{\infty} \frac{\sigma_r(\omega')}{\omega'^2 - \omega^2}d\omega'\,,
\end{equation}
where $\mathcal{P}$ denotes the principal value of the integral. 

\begin{figure} [ht]
   \begin{center}
   \setlength\tabcolsep{0pt} 
   \includegraphics[width=3.3in]{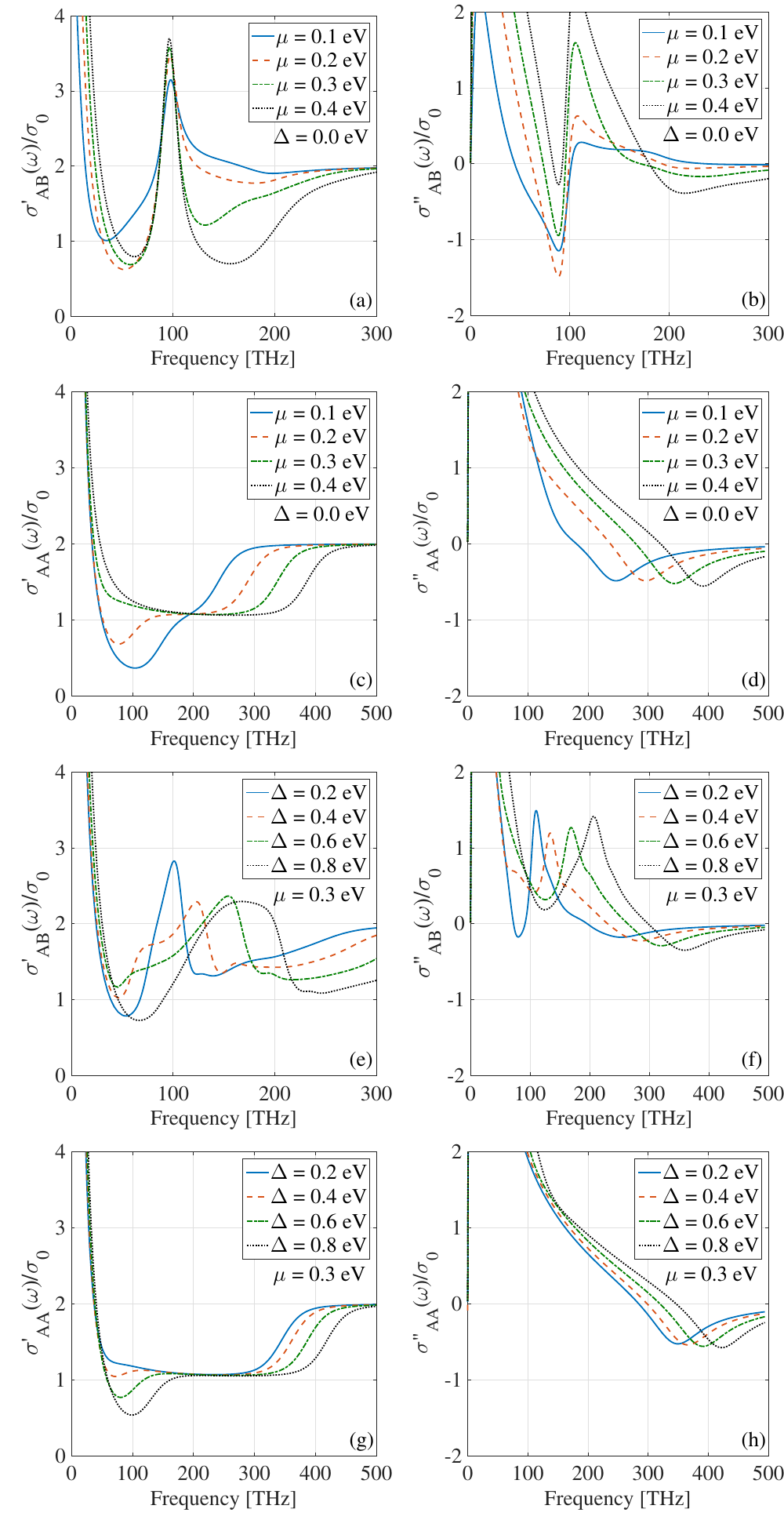}
   \end{center}
   \caption[example] 
   { \label{fig:oc} 
Optical conductivity of BLG, for different values of Fermi level $\mu$ and energy asymmetry $\Delta$, numerically evaluated by using Eqs. (\ref{eq:oc-ab-sigma}), (\ref{eq:oc-aa-sigma}) and (\ref{eq:oc-imag}), normalized to $\sigma_0 = q^2/4\hbar$. The figures illustrate the impact of $\mu$ on the optical conductivity of (a, b) AB stacked BLG and (c, d) AA stacked BLG for a fixed $\Delta$ and the impact of $\Delta$ on the optical conductivity of (e, f) AB stacked  BLG and (g, h) AA stacked BLG for a fixed $\mu$. All simulations are conducted at $300$ K with a constant scattering rate of $\Gamma=20$ meV.}
   \end{figure}

Figure \ref{fig:oc} shows the optical conductivity of AB and AA stacked BLG for different values of $\mu$ and $\Delta$ using an energy-independent value of electron scattering rate $\Gamma=20$ meV at room temperature $T = 300$ K.
The plots are normalized to the universal conductivity of graphene $\sigma_0 = q^2/4\hbar$. 
In Fig.~\ref{fig:oc} (a, c), the real parts of the optical conductivity ($\sigma'(\omega)$) of AB and AA stacked BLG are illustrated for different values of the Fermi level, $\mu$ at $\Delta=0$ eV.
We can see that for a fixed energy asymmetry, the optical conductivity can be tuned by the Fermi level. For instance in the case of AA stacked BLG, the interband transition is shifted in frequency as the Fermi level changes. 
In the case of AB stacked BLG, there is a strong absorption peak at frequencies around $\gamma_1$ due to interband transitions. 
Figures \ref{fig:oc} (e, g) illustrate $\sigma'(\omega)$ for different values of $\Delta$ at a fixed Fermi level $\mu=0.3$ eV. 
From these plots we can see that the impact of energy asymmetry on the optical conductivity of BLG is quite significant specially in the case of AB stacked BLG where the absorption peak shifts toward higher frequencies as $\Delta$ increases. 
This phenomenon can be seen as an additional degree of freedom for tuning the optical conductivity of BLG that is absent in the case of MLG.
At frequencies close to the visible region, the real part of conductivity converges to $2\sigma_0$ as expected\cite{nair2008fine}.
The imaginary parts of the optical conductivity $\sigma''(\omega)$ are plotted in Figs. \ref{fig:oc} (b, d, f, h). 
These figures show the existence of different frequency bands where TE plasmons ($\sigma''(\omega)<0$) and TM plasmons ($\sigma''(\omega)>0$) can be supported and tuned in AA and AB stacked BLG structures.

\subsection{\label{sec:gate} Gate Voltage and Electrostatic Screening}
\begin{figure} [ht]
   \begin{center}
   \includegraphics[width=2.8in]{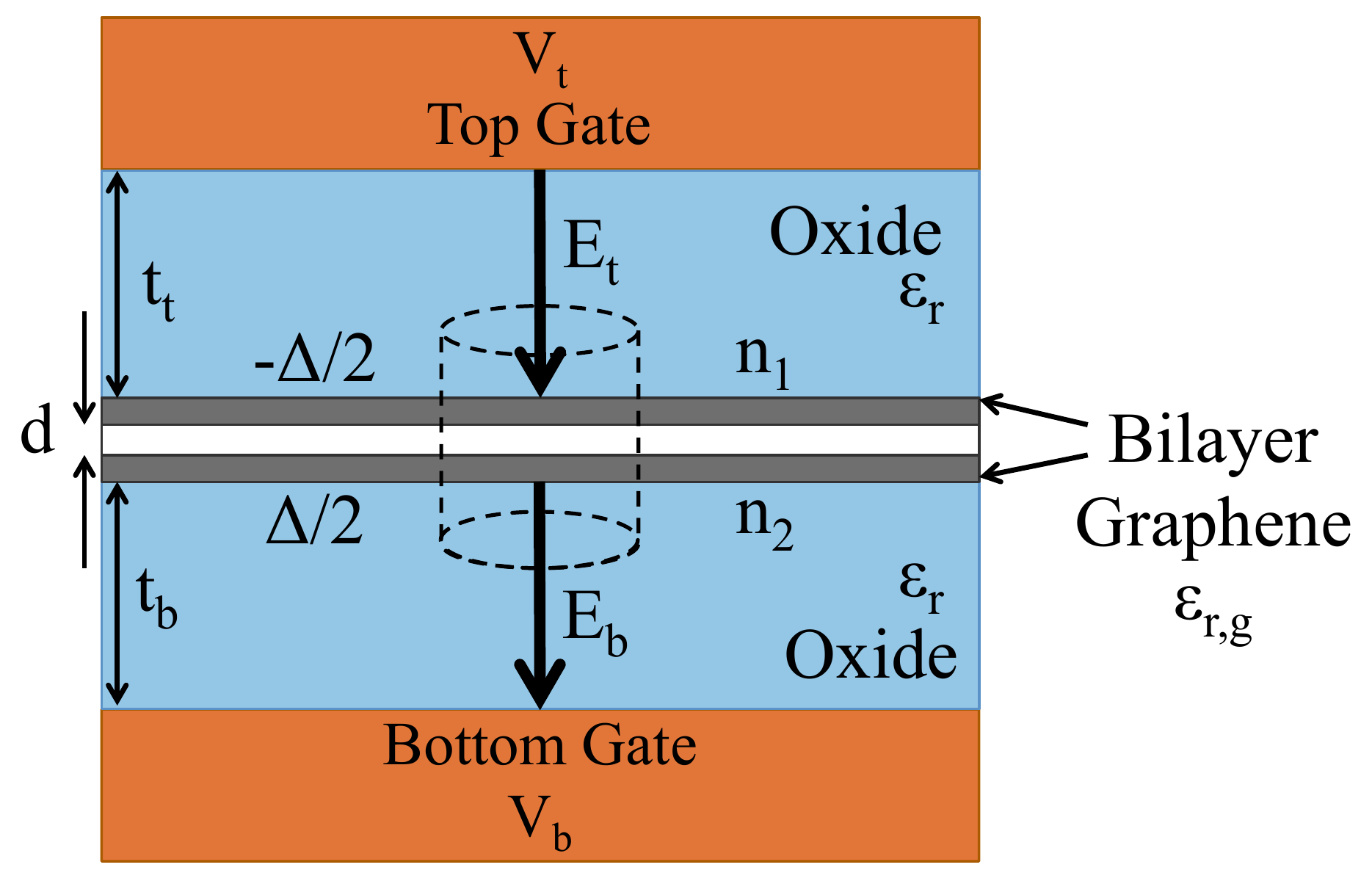}
   \end{center}
   \caption[example] 
   { \label{fig:gate} 
Gated BLG within a uniform dielectric medium with relative permittivity of $\varepsilon_r$. Gate oxide thickness is denoted as $t_{t}$ for the top oxide and $t_b$ for the bottom oxide. The interlayer distance in BLG is denoted as $d=0.35$ nm. Dual gate structure is used to control Fermi level $\mu$ and energy asymmetry $\Delta$ independently. The carrier density on graphene layers are represented by $n_1$ and $n_2$. The difference between oxide fields $\delta E = E_b - E_t$ determines the Fermi level and their average $\overline{E} = (E_b + E_t)/2$ introduces the energy asymmetry. We assume that the relative permittivity of BLG is equal to $\varepsilon_{r,g}=1$.}
\end{figure}

Experimental observations \cite{zhang2009direct, oostinga2008gate, ohta2006controlling, castro2007biased} as well as theoretical studies \cite{mccann2006asymmetry, mccann2007low, castro2008bilayer, nilsson2007transmission, min2007ab, avetisyan2009electric} have demonstrated that applying an electrostatic field perpendicular to the BLG plane forms an energy asymmetry between the two layers of graphene and consequently alters its bandstructure. 
In Sec. \ref{sec:band} and \ref{sec:oc}, we discussed that
the presence of energy asymmetry creates a bandgap in the bandstructure of AB stacked BLG, while it modifies the interlayer hopping energy $\gamma_1$ of AA stacked BLG. These phenomena introduce an additional degree of freedom (along with the Fermi level) for tuning the optical conductivity of BLG. This eletctric-field-induced energy asymmetry can be realized by gating the graphene layers as shown in Fig.~\ref{fig:gate}. We use the dual gate structure shown in this figure to obtain the energy asymmetry, $\Delta$, in terms of top and bottom gate voltages ($V_t$ and $V_b$ respectively).
In this metal-oxide-graphene geometry, the gate voltage applied at the metal electrode, modulates the energy asymmetry and the carrier density in BLG via the electrostatic field in the oxide. The physical thickness of the oxide is denoted as $t_{t}$ for the top oxide and $t_b$ for the bottom oxide.
The parameter $d= 0.35$ nm is the interlayer distance of BLG and is equal to the thickness of each layer. 
The use of dual gates allows us to independently control the Fermi level and energy asymmetry in the BLG.
The fields of top and bottom gates are $E_t$ and $E_b$, respectively. 
It has been shown that the difference between these two fields $\delta E = E_b - E_t$ determines the Fermi level, and their average $\overline{E}=(E_b + E_t)/2$ introduces the energy asymmetry\cite{zhang2009direct}. 
For given gate voltages, $V_t>0$ and $V_b<0$, by utilizing Gauss's law, we derive the total carrier density induced on the graphene layers as
\begin{equation}
\label{eq:n-voltage}
n = n_t + n_b  = -\frac{\varepsilon_r\varepsilon_0}{q} \delta{E} = \frac{\varepsilon_r\varepsilon_0}{qt_{t}}V_t + \frac{\varepsilon_r\varepsilon_0}{qt_b}V_b\cdot
\end{equation}
Here $n_t$ and $n_b$ are the carrier densities induced by the the top and bottom gates respectively and $q$ is the magnitude of the electron charge. We note that for the negative charged carriers, $n$ is positive. The Fermi level is related to the total carrier density by the following equation 
 \begin{equation}
\label{eq:n-fermi}
\begin{split}
n &= \sum_\alpha n_\text{electrons} - n_\text{holes} \\&= \sum_\alpha \int_{0}^{\infty} \rho_\alpha(E)f(E)dE - \int_{-\infty}^{0} \rho_\alpha(E)[1 - f(E)]dE\, ,
\end{split}
\end{equation} 
where $\rho_\alpha(E)$ is the density of states of the $\alpha^\text{th}$ subband and is defined as follows \cite{ashcroft1976introduction}
 \begin{equation}
\label{eq:dos}
\rho_\alpha(E) = 4\int\frac{d^2{\bm k}}{(2\pi)^2}\delta(E - \epsilon_\alpha({\bm k}))\cdot
\end{equation} 
 \begin{figure} [ht]
   \begin{center}
   \includegraphics[width=2.8in]{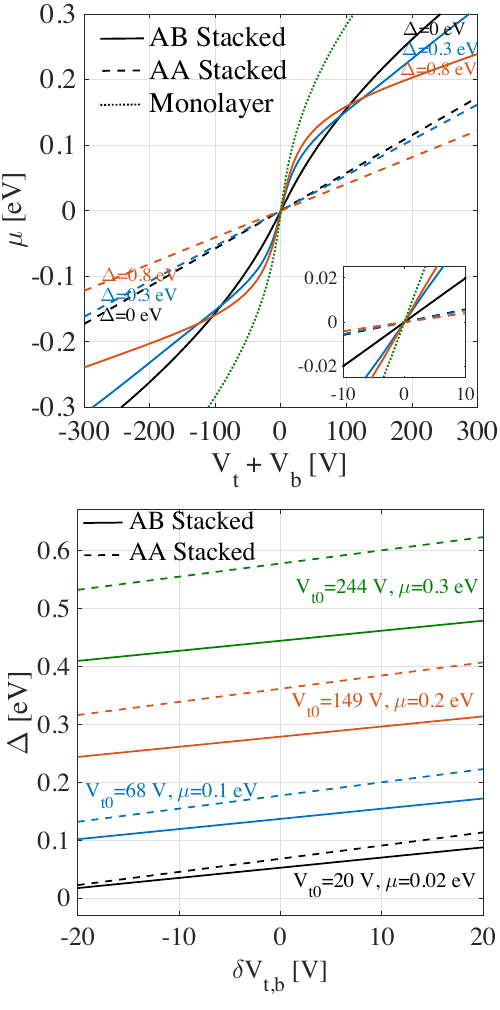}
   \end{center}
   \caption[example] 
   { \label{fig:control-fermi-delta} 
Impact of gate on the Fermi level $\mu$ (top) and the energy asymmetry $\Delta$ (bottom) in BLG. The total gate voltage, $V_t + V_b$, determines $\mu$ and the differential gate voltage $\delta V_{t,b}$ determines $\Delta$. The inset shows the curves near zero gate voltage. The energy asymmetry is calculated in the presence of imperfect electrostatic screening. In the bottom figure the initial top gate voltage is denoted by $V_{t0}$. The initial bottom gate voltage, $V_{b0}=-10$ V, is the same for all the cases.}
\end{figure} 
We have included the valley and spin degeneracies using a factor of four in the above equation. Using Eqs. (\ref{eq:n-voltage}) and (\ref{eq:n-fermi}), we find the Fermi level as a function of the top and bottom gate voltages. 
Figure \ref{fig:control-fermi-delta} shows how the Fermi level of BLG varies with respect to $V_t + V_b = t_{t}\delta E$ for a fixed $\Delta$. 
Here we have assumed that top and bottom oxides are made of SiO$_2$ with permittivity $\varepsilon_r = 3.9$ and that they have equal thicknesses $t_t=t_b=300$ nm. 
As we can see from the plots, the Fermi level of monolayer and AB stacked BLG show a higher sensitivity to the gate voltage than the Fermi level of AA stacked BLG. Also, at high gate voltages, different energy asymmetries lead to different curves. 
The inset illustrates the curves for gate voltages close to zero. 
The average electric field induced by the gate contacts ($\overline{E}=(V_t - V_b)/2t_t$) polarizes the graphene layers, which leads to an internal electric field given as $E_{int} = -q\delta n/2\varepsilon_{r,g}\varepsilon_0$ with $\delta n$ as the charge imbalance between the two layers and $\varepsilon_{r,g}=1$ as the relative permittivity of BLG. The internal electric field acts to cancel out the external electric field.
This effect, known as electrostatic screening, plays a major role in realistic multilayer graphene devices and is carefully modeled in our analysis.
Due to the finite electrostatic screening in BLG, the electron density on the top layer $n_1=(n +\delta n)/2$ is not equal to the electron density on the bottom layer $n_2=(n -\delta n)/2 $. 
We note that the charge conservation is satisfied, $n_1 + n_2 = n_t + n_b$.
The charge imbalance $\delta n = n_1 - n_2$ is given as the relative weights of electron wavefunctions as follows \cite{mccann2007low}
\begin{equation}
\label{eq:charge-imbalance}
\begin{split}
\delta n = n_1 - n_2 = \frac{2}{\pi}\int \sum_{\alpha} (|\psi^\alpha_{A1}(k)|^2 + |\psi^\alpha_{B1}(k)|^2 \\-|\psi^\alpha_{A2}(k)|^2 - |\psi^\alpha_{B1}(k)|^2)k\,dk,
\end{split}
\end{equation}
where $\big(\psi^\alpha_{A1}(k), \psi^\alpha_{B2}(k), \psi^\alpha_{A2}(k), \psi^\alpha_{B1}(k)\big)$ represents the $\alpha^\text{th}$ eigenvector of the Hamiltonian in Eqs. (\ref{eq:band-ab-hamil}) and (\ref{eq:band-aa-hamil}), and the integral is evaluated over all occupied states. Therefore, we can write the energy asymmetry as 
\begin{equation}
\label{eq:self-consistent}
\Delta = q(\overline{E} + E_{int})d = (V_t - V_b)qd/2t_t - \delta n(\Delta)\frac{q^2d}{2\varepsilon_r\varepsilon_0}\, ,
\end{equation}
which is the same as the self-consistent Hartree approximation in Ref. \onlinecite{castro2008bilayer}. 
By setting $V_t + V_b$ to a constant value and applying a differential voltage $\delta V_{t,b}$ to the gate contacts ($V_t$ becomes $V_t + \delta V_{t,b}/2$ and $V_b$ becomes $V_b - \delta V_{t,b}/2$), we fix the Fermi level and can independently study the impact of gate voltage on the energy asymmetry.
Figure \ref{fig:control-fermi-delta} illustrates the energy asymmetry as a function of differential gate voltage, $\delta V_{t,b}$, for a variety of fixed Fermi levels. As we can see the energy asymmetry has a linear dependence on $\delta V_{t,b}$. Moreover, for the given gate voltages, AA stacked BLG shows a higher energy asymmetry than AB stacked BLG. We note that the curves are also valid for larger differential gate voltages as long as the oxide field is below the dielectric breakdown strength of SiO$_2$, which is larger than $1$ V/nm.

\section{\label{sec:results} Results}
In this section, we discuss the features of propagating TM and TE plasmons, such as compression factor, mode confinement, and propagation length, in various graphene structures. 
From the perspective of graphene plasmonic devices (antennas and waveguides), it is desirable to simultaneously achieve confined (low transverse propagation length $1/2k''_x$) waves with a low loss (high longitudinal propagation length $1/2k''_z$) and a high compression factor $k'_z/k_\text{light}$. 
We use the complex wavevector derived in Sec. \ref{sec:plasmon} and combine it with the conductivity model from Sec. \ref{sec:oc} in the presence of gate effects (Sec. \ref{sec:gate}) for analysis in this section.
\begin{figure} [ht]
   \begin{center}
   \includegraphics[width=3.3in]{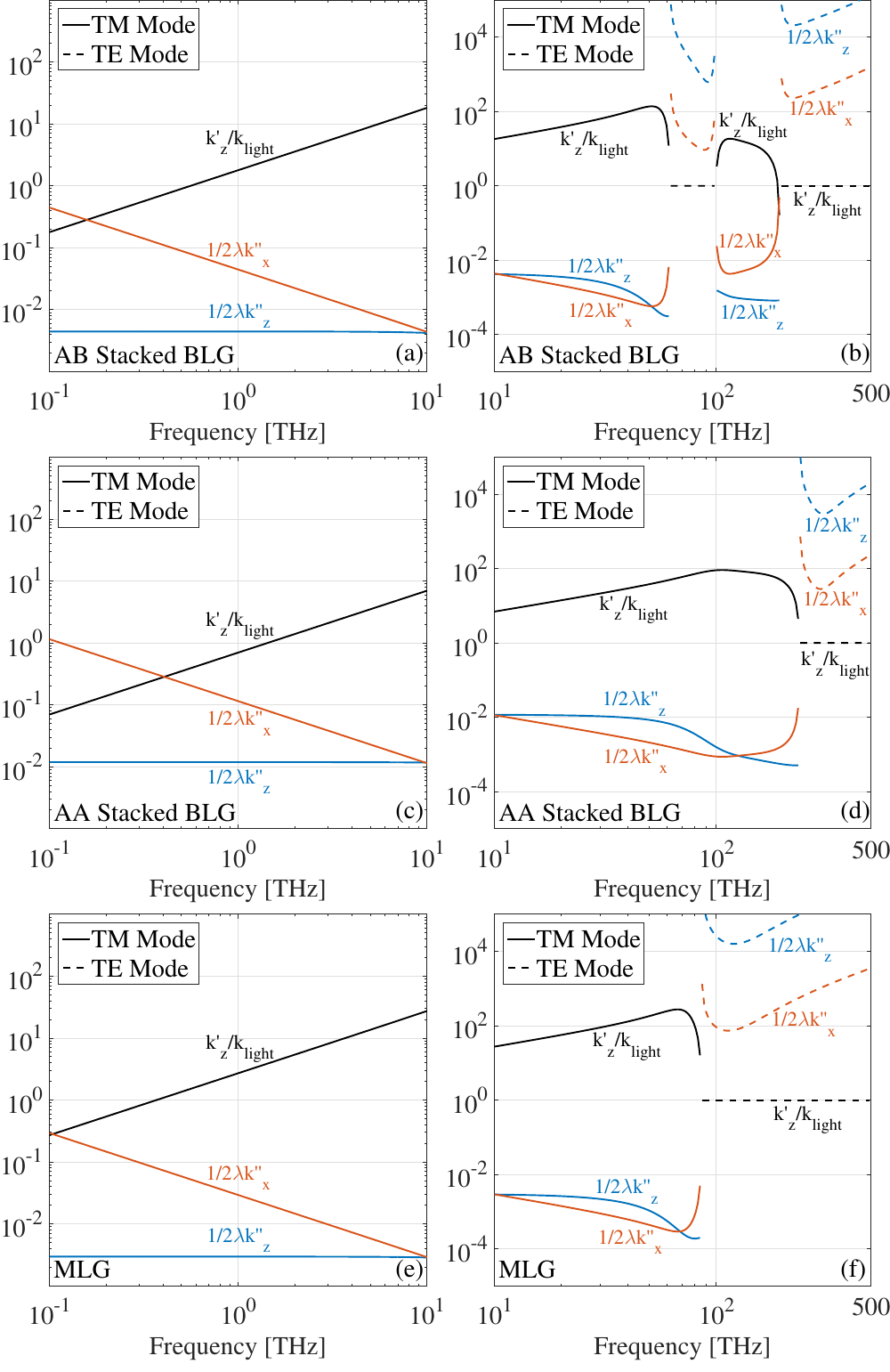}
   \end{center}
   \caption[example] 
   { \label{fig:plasmon-freq} 
Relative phase constant $k'_z/k_\text{light}$, longitudinal propagation length $1/2\lambda_zk''_z$, and transverse propagation length $1/2\lambda_xk''_x$ as a function of frequency for AB stacked BLG (a, b), AA stacked BLG (c, d), and MLG (e, f) in the terahertz (left column) and infrared (right column) regimes. Here we assume that the Fermi level is $\mu=0.3$ eV, energy asymmetry is $\Delta=0$ eV. The wavevectors are calculated using Eqs. (\ref{eq:dispersion-te-detailed}) and (\ref{eq:dispersion-tm-detailed}).}
   \end{figure} 
   
In Fig. \ref{fig:plasmon-freq}, we plot the following plasmon characteristics versus the operating frequency: (i) relative longitudinal propagation length $1/2\lambda k''_z$ normalized to the wavelength of light $\lambda=2\pi/k_\text{light}$, (ii) relative phase constant $k_z'/k_\text{light}$ normalized to $k_\text{light}$, and (iii) relative transverse propagation constant $1/2\lambda k''_x$ normalized to $\lambda$.  
The results in this figure are obtained for AB and AA stacked BLG, and MLG in the terahertz and infrared regimes for $\mu=0.3$ eV and $\Delta = 0$ eV. 
The conductivity of MLG is derived using Green's functions approach described for BLG in Sec. \ref{sec:oc}. It can  also be derived using the procedure outlined in Ref. \onlinecite{falkovsky2007optical}.
In the terahertz frequency band (left column of Fig. \ref{fig:plasmon-freq}), all the graphene structures support TM mode propagation with similar propagation characteristics. As the frequency increases the plasmon waves become more compressed and more confined to the graphene sheet.
While MLG has a relatively higher compression factor, its propagation length is smaller than that of BLG. 
In the infrared regime (right column of Fig. \ref{fig:plasmon-freq}), there exists separate frequency bands that allow the propagation of both TM and TE modes in graphene structures.
Generally, TE mode in both BLG and MLG has a phase constant ($k'_z$) very close to the phase constant of light ($k_{light}$) and shows a very high propagation length in longitudinal direction but they are very loosely bounded to the graphene sheet. Conversely, TM mode mostly shows a high compression factor, but very low propagation length which is two orders of magnitude lower than the wavelength of light. However, they are more confined in the transverse direction than TE modes and so are better guided by the graphene structure. 
The trade-off between the highly confined TM mode and the highly propagating TE mode is a general characteristic of plasmons and has been reported previously\cite{hanson2008dyadic, jablan2009plasmonics, maier2007plasmonics}. 
The reason is that high confinement to the graphene sheet implies that a large amount of plasmon wave energy is 
confined to the sheet which has a higher loss than the surrounding dielectric medium.  

\begin{figure} [ht]
   \begin{center}
   	\includegraphics[width=3.5in]{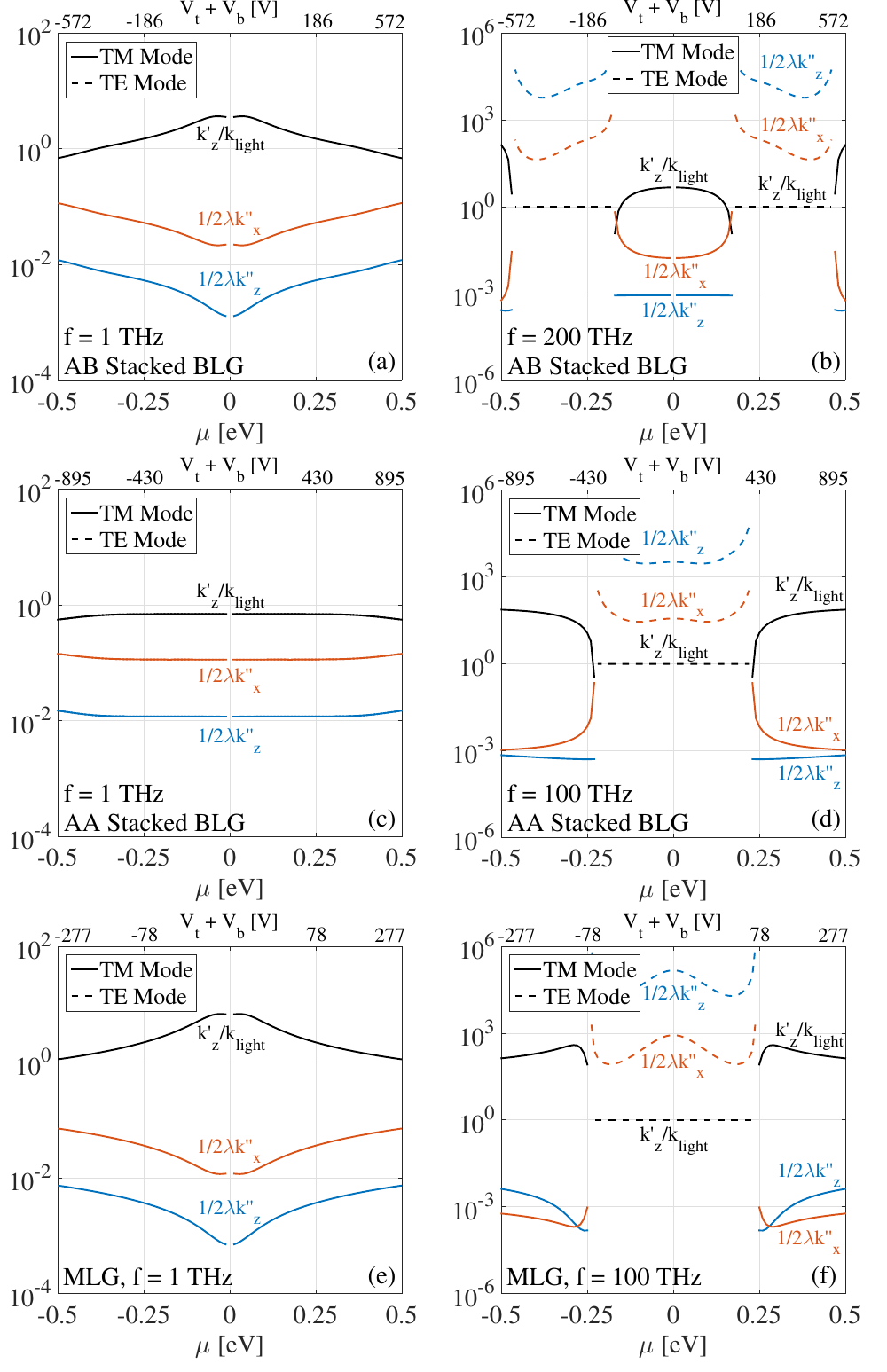}
   \end{center}
   \caption[example] 
   { \label{fig:plasmon-ef} 
Impact of Fermi level $\mu$ on the relative phase constant $k'_z/k_\text{light}$, longitudinal propagation length $1/2\lambda_zk''_z$, and transverse propagation length $1/2\lambda_xk''_x$ for AB stacked BLG (a, b), AA stacked BLG (c, d), and MLG (e, f) for a given frequency in the terahertz band (left column) and infrared band (right column). The total gate voltages corresponding to each Fermi level is labeled on the top axis. The wavevectors are calculated using Eqs. (\ref{eq:dispersion-te-detailed}) and (\ref{eq:dispersion-tm-detailed}).} 
   \end{figure}

To study the impact of chemical potential, we plot the various plasmon characteristics in graphene structures as a function of the chemical potential $\mu$ for a constant frequency in the terahertz band and a constant frequency in the infrared band in Fig.~\ref{fig:plasmon-ef}. 
The gate voltages corresponding to the Fermi levels are labeled on the top axis. We note that very high Fermi levels are not achievable in practice by mere gating since it would require electric fields higher than the break down field of SiO$_2$. 
In our device setup, oxide voltages cannot exceed $300$ V, however to obtain high Fermi levels, one can utilize chemical doping which would basically act as a gate with a constant induced carrier density.  
For the $1$ THz frequency (left column of Fig. \ref{fig:plasmon-ef}), it is evident that as the Fermi level increases, AB stacked BLG and MLG show an increase in both longitudinal and transverse propagation lengths and a decrease in compression factor. However, plasmons in AA stacked BLG show little sensitivity to changes in the Fermi level.
Moreover, similar to the trade-off between TM and TE modes, we observe a trade-off between compression factor and propagation length of TM mode plasmons as we chcange the Fermi level. 
At infrared frequencies (right column of Fig. \ref{fig:plasmon-ef}), by changing the Fermi level, we can switch the mode of propagation between the highly confined TM mode and the loosely bounded TE mode.

\begin{figure} [ht]
   \begin{center}
   \includegraphics[width=3.4in]{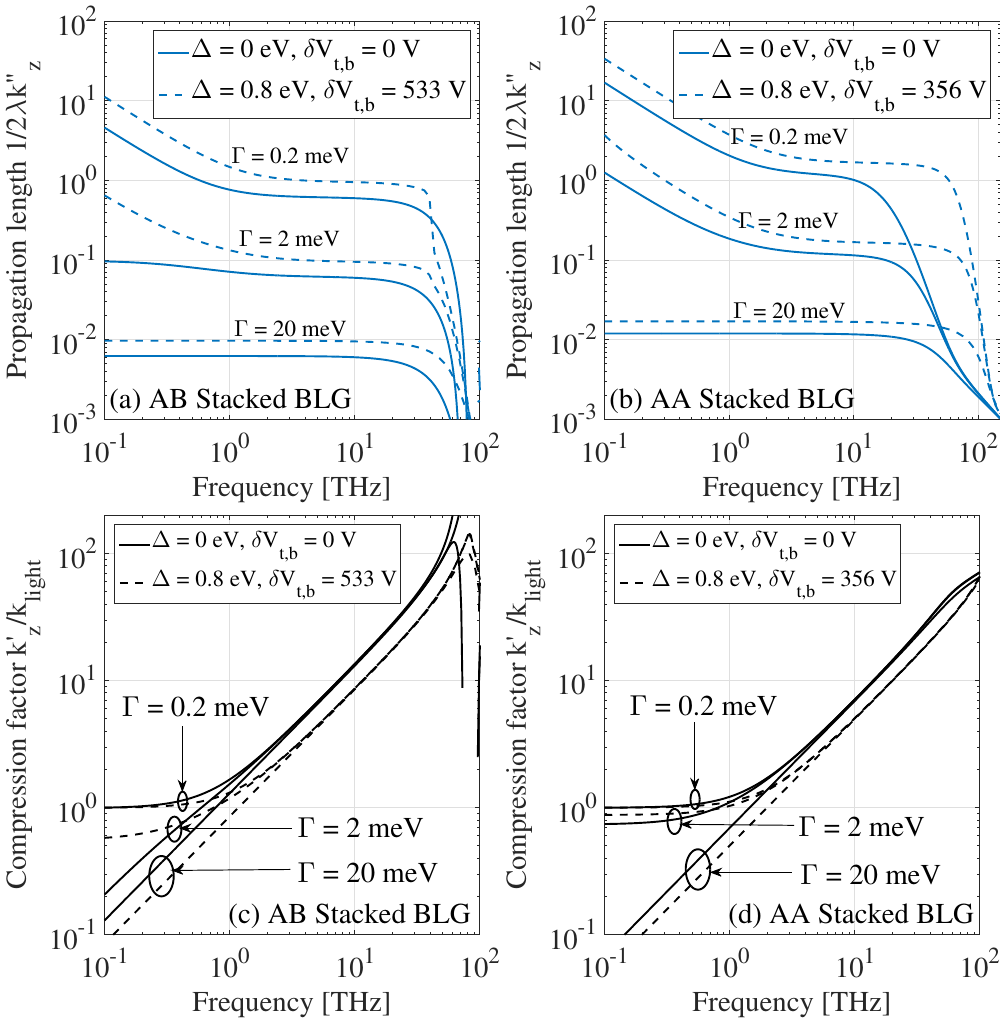}
   \end{center}
   \caption[example] 
   { \label{fig:plasmon-del} 
Impact of Energy asymmetry $\Delta$ and scattering rate $\Gamma$ on the longitudinal propagation length $1/2\lambda_zk''_z$ (a, b) and relative phase constant $k'_z/k_\text{light}$ (c, d) for AB and AA stacked BLG. The propagation length can be improved by increasing $\Delta$ but at the expense of compression factor. The differential gate voltage required to create $\Delta$ is denoted as $\delta V_{t,b}$. Here we assume that the Fermi level is $\mu=0.3$ eV. The wavevectors are calculated using Eqs. (\ref{eq:dispersion-te-detailed}) and (\ref{eq:dispersion-tm-detailed}).}
   \end{figure}

In Sec. \ref{sec:oc} we discussed that in addition to the Fermi level, the energy asymmetry in BLG can be utilized as an extra degree of freedom to tune its optical conductivity. To study the impact of energy asymmetry, we plot the propagation characteristics of plasmons in BLG versus frequency for two values of energy asymmetry ($\Delta=0$ and $0.8$ eV) and a fixed Fermi level in Fig. \ref{fig:plasmon-del}.
The Fermi level needs to be sufficiently far from the charge neutrality point so that there are enough carriers to enable plasmon oscillations. Specifically, due to the induced bandgap in AB stacked BLG, there exists a plasmon cut-off region ($\mu < 0.1$ eV) where the carrier concentration is not sufficient for plasmon oscillations.\cite{fei2015tunneling, chen2012optical, woessner2015highly}. 
Here, the Fermi level is fixed at $0.3$ eV.
Each curve in Figs. \ref{fig:plasmon-del}a and \ref{fig:plasmon-del}b corresponds to a different value of scattering rate for both $\Delta=0$ eV (solid) and $\Delta=0.8$ eV (dashed). 
As we see from the figure, scattering rate has an enormous impact on plasmon waves decay. Hence, the ability to produce high quality samples with low impurity and defect concentrations becomes an essential challenge of building high performance graphene based plasmonic devices.

Interestingly, the propagation length of plasmons in AB and AA stacked BLG can be improved by increasing $\Delta$. 
Plasmons in AB stacked BLG can propagate 2 to 5 times longer, depending on the frequency band and the scattering rate. 
Similarly, plasmons in AA stacked BLG can propagate 1.5 to 100 times longer.
The improvement in propagation length is mainly due to the reduction in absorption ($\sigma'(\omega)$) caused by supression of interband transitions such as second valance subband to second conduction subband ($\epsilon_{-,2}\rightarrow\epsilon_{+,2}$) and first conduction subband to second conduction subband ($\epsilon_{+,1}\rightarrow\epsilon_{+,2}$). 
This a significant benefit of BLG as it will enable low energy and low leakage plasmonic antennas and waveguides for THz and low infrared bands communication. 
However, these improvements come at the expense of smaller compression factor (Fig. \ref{fig:plasmon-del}c, d) as mentioned earlier. 

\begin{figure} [ht]
   \begin{center}
   \includegraphics[width=3.4in]{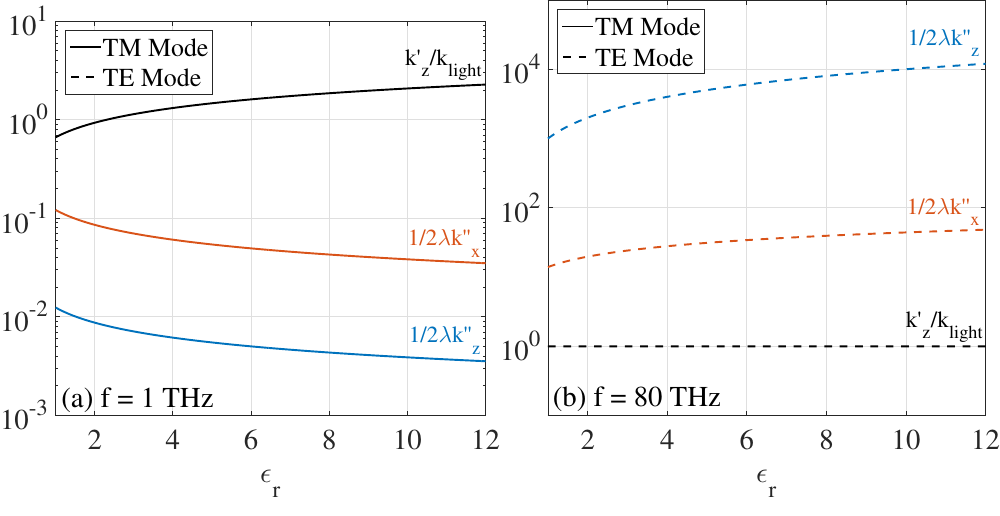}
   \end{center}
   \caption[example] 
   { \label{fig:plasmon-eps} 
Relative phase constant $k'_z/k_\text{light}$, longitudinal propagation length $1/2\lambda_zk''_z$, and transverse propagation length $1/2\lambda_xk''_x$ as a function of dielectric constant $\varepsilon_r$ for AB stacked BLG for TM ($1$ THz - solid) and TE ($80$ THz - dashed) modes. Here we assume that the Fermi level is $\mu=0.3$ eV, $\Gamma=20$ meV, and $T=300$ K. The wavevectors are calculated using Eqs. (\ref{eq:dispersion-te-detailed}) and (\ref{eq:dispersion-tm-detailed}).}    
\end{figure}

Finally we illustrate how the plasmon characteristics are affected by the medium dielectric. Figure \ref{fig:plasmon-eps} illustrates the plasmon characteristics as a function of the relative permittivity of dielectric medium for TM and TE modes of propagation.  
As we can see, while increasing dielectric constant, lowers propagation length of the TM mode, it increases propagation length of the TE mode. This effect can also be seen directly from Eqs. (\ref{eq:dispersion-te-detailed}) and (\ref{eq:dispersion-tm-detailed}) where the dependence on dielectric constant appears in $\eta$. Moreover, higher compression factors can be achieved for TM mode by increasing dielectric constant. 

\section{\label{sec:conclusion} Conclusions}
The focus of this paper is to quantify the impact of material parameters on the tunability of plasmon propagation characteristics in bilayer graphene (BLG) structures. 
Toward this end, we consider a dual gate metal-oxide-graphene geometry to independently tweak the Fermi level and energy asymmetry in BLG through the gate terminal voltages. 
We numerically obtain the relationship between energy asymmetry and the applied gate voltage using the self-consistent Hartree approximation. 
The optical conductivity of BLG is modeled as a function of Fermi level and energy asymmetry using the Kubo formalism. 
The propagation of surface plasmons is studied numerically to quantify their propagation length, mode confinement, and compression factor in various graphene structures including AB stacked and AA stacked BLG, and monolayer graphene (MLG). 
Important trade-offs between propagation length and mode confinement that must be considered for device design are discussed. 
Impact of scattering rate on the propagation length of plasmons is also quantified. 
Finally, we demonstrate that the propagation length of TM plasmons in BLG can be improved by increasing the energy asymmetry. 
Such an advantage is absent in MLG. 
Overall, the results presented in this work provide insight into the design and optimization of plasmonic structures utilizing graphene as their platform.


\appendix
\section{\label{app:scattering}  Impact of energy asymmetry on scattering rate}
\setcounter{figure}{0}
\renewcommand{\thefigure}{A\arabic{figure}}
In general, the scattering rate depends on the bandstructure and the Fermi velocity. A larger energy asymmetry would reduce the Fermi velocity of electrons. 
However, Fermi velocity is not the only factor that affects scattering rate and correspondingly the various figures of merit of plasmons.
Factors including polarizability and screening, which depend on $\Delta$, are important to consider, particularly in the case of charged impurity scatterers.
These factors could outweigh the reduction of Fermi velocity and even decrease the scattering rate.
To study the nontrivial dependence of scattering rate on $\Delta$, we numerically calculate the average scattering rate, $\Gamma = 1/2\braket{\tau}$, where $\braket{\tau}$ is the average carrier relaxation time and is given as \cite{hwang2008acoustic}
\begin{equation}
\label{eq:tau-avg}
\braket{\tau} = \frac{\sum_{s,\alpha}\int \rho_{s,\alpha}(E)\tau_{s,\alpha}(E)(-\frac{df(E)}{dE})dE}
{\sum_{s,\alpha}\int \rho_{s,\alpha}(E)(-\frac{df(E)}{dE})dE}\,\cdot
\end{equation}
Here, $\tau_{s,\alpha}(E)$ is the energy-dependent relaxation time corresponding to the band $s$ and subband $\alpha$. Density of states is denoted as $\rho_{s, \alpha}(E)$, and $f(E)$ is the Fermi-Dirac distribution function.
The k-dependent relaxation time $\tau_{s,\alpha}({\bm k})=\tau_{s, \alpha}(E)|_{E = \epsilon_{s, \alpha}({\bm k})}$ is written using Boltzmann approximation as follows\cite{sarma2010theory}
\begin{equation}
\label{eq:tau-k}
\begin{split}
\frac{1}{\tau_{s,\alpha}({\bm k})} = \frac{2\pi n_0}{\hbar}\int \frac{d^2{\bm k'}}{(2\pi)^2} |\braket{V_{s, \alpha}({\bm k, \bm k'})}|^2 \times \\
(1 - \cos(\theta_{\bm k, \bm k'})) \delta(\epsilon({\bm k}) - \epsilon({\bm k'}))\, ,
\end{split}
\end{equation}
where $n_0$ is the density of scatterers, $\theta_{\bf k, k'}$ is the angle between the incident wavevector ${\bf k}$ and the scattered wavevector ${\bf k'}$, and $\braket{V_{s, \alpha}({\bf k, k'})}$ is the matrix element of the disorder potential. 
$\braket{V_{s, \alpha}({\bf k, k'})}$ depends on the scattering mechanism and the electron wavefunctions $\psi_{s, \alpha}({\bm k})$ and is written as 
\begin{equation}
\label{eq:potential}
\braket{V_{s, \alpha}({\bm k, \bm k'})} =  
\psi_{s, \alpha}^{\dagger}({\bm k})\psi_{s, \alpha}({\bm k'})
V({\bm q})\,,
\end{equation}
Here, ${\bm q}={\bm k'}-{\bm k}$ and $V({\bm q})=\int  V({\bf r})e^{i{\bf q}\cdot{\bf r}}  d{\bf r}$ is the Fourier transform of $V(\bm r)$, which is the scattering potential.
\begin{figure} [ht]
\begin{center}
\includegraphics[width=2.5in]{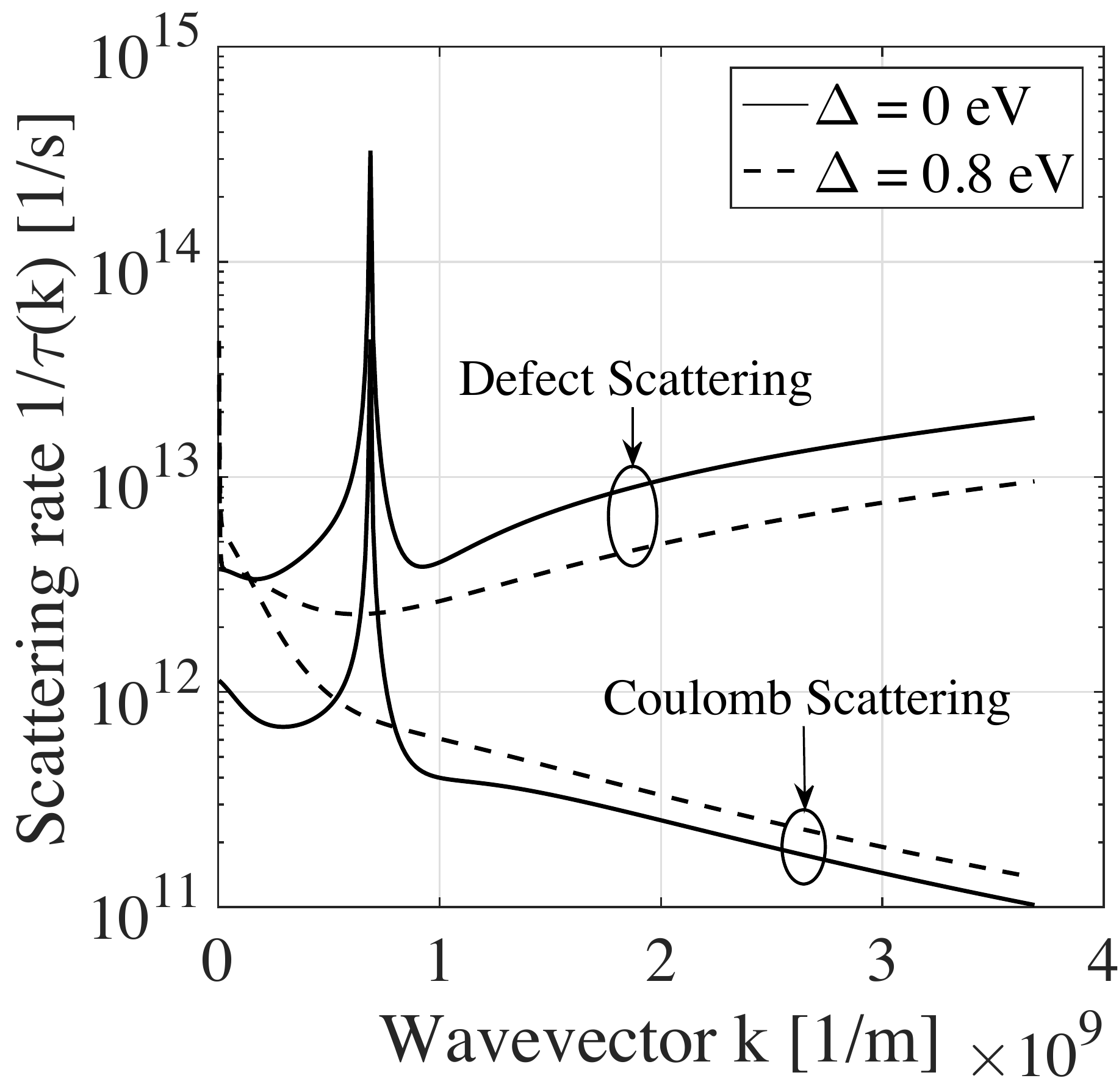} 
\end{center}
\caption
{Coulomb and defect scattering rates as functions of wavevector for two values of energy asymmetry i.e. $\Delta=0$ and $\Delta=0.8\,$eV evaluated using Eq. \ref{eq:tau-k}. The concentration of Coulomb scatterers is $n_i=1\times10^{16}\,$m$^{-2}$ and the concentration of defects is $n_d=0.01\times10^{16}\,$m$^{-2}$. The dielectric media is $\text{SiO}_2$ with relative permittivity of 3.9.
} 
\label{fig:tau} 
\end{figure}
We consider long-range Coulomb scattering and short-range defect scattering mechanisms.
In the presence of screening, for charged impurities located at distance $d$ from the BLG sheet we have $V({\bm q})=v(q)\exp(-qd)/\varepsilon(q)$, where $v(q)=e^2/2\varepsilon_{r}\varepsilon_{0}q$ is the bare coulomb potential.
The dielectric function is $\varepsilon(q)=1 + v(q)\Pi(q)$, and $\Pi(q)$ represents the polarizability of BLG. The details of calculation of polarizability are provided in Refs. \onlinecite{hwang2008screening} and \onlinecite{lv2010screening}. For defect scattering $V({\bm r})$ is a delta function in space; therefore, its Fourier transform is constant and independent of the dielectric function and polarizability i.e. $V({\bm q})=V_0$. Here we assume that $V_0=10\,$eVnm$^{2}$(Ref. \onlinecite{sarma2011electronic}).
Figure \ref{fig:tau} illustrates k-dependent scattering rates, evaluated numerically using Eq. \ref{eq:tau-k}, for both coulomb and defect scattering mechanisms, and for two values of energy asymmetry (i.e. $\Delta=0$ and $\Delta=0.8\,$eV). Material-specific parameters used to obtain the results reported in Fig.~\ref{fig:tau} are provided in the figure caption. The figure highlights the opposite behaviors of Coulomb and defect scattering in response to introducing a large $\Delta$. For Coulomb scattering, scattering rate decreases with an increase in momentum\cite{monteverde2010transport} because of the increase in screening. Due to the absence of screening in the case of defect scattering, scattering rate increases with wavevector. 

\begin{figure} [ht]
\begin{center}
\includegraphics[width=3.4in]{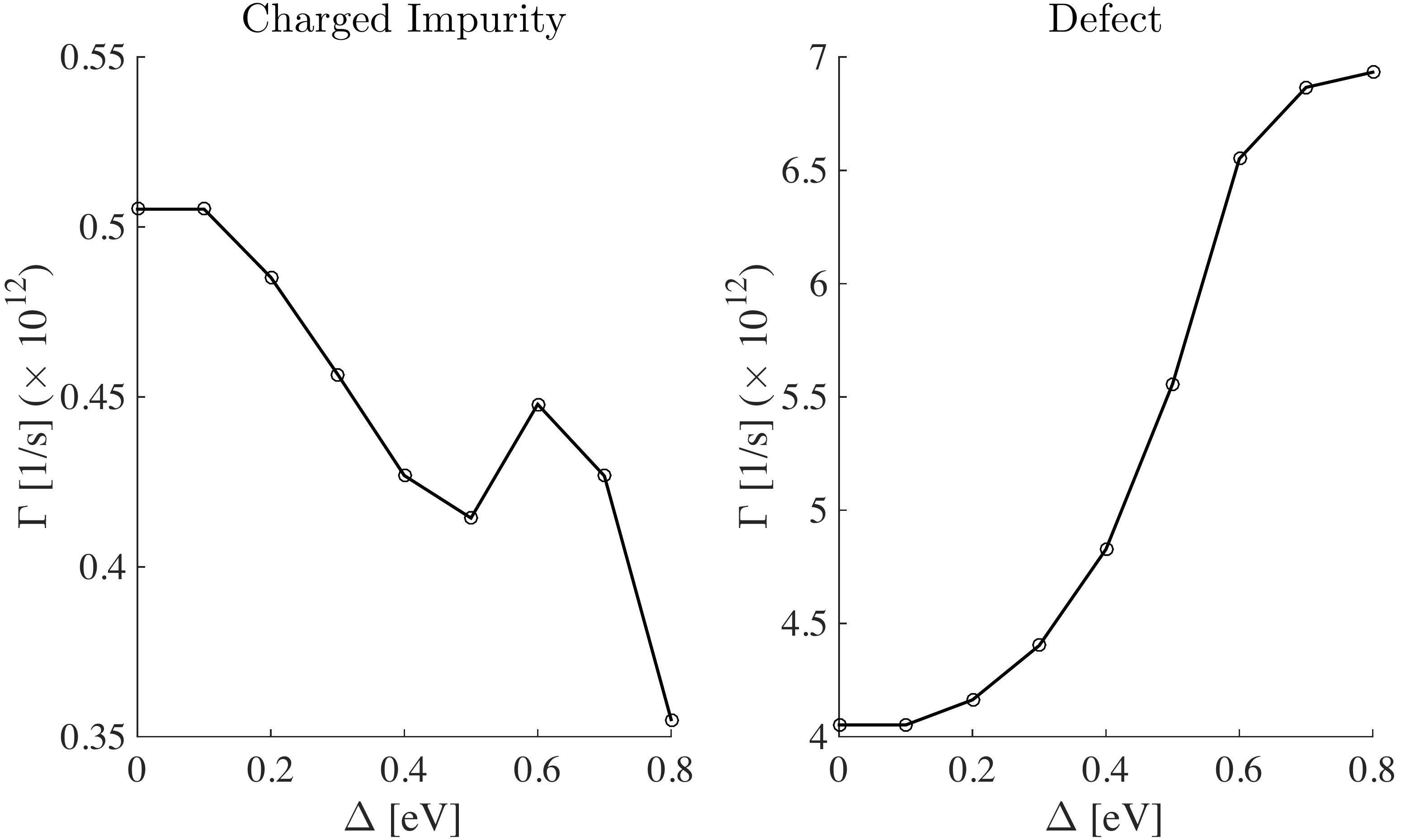} 
\end{center}
\caption 
{Average scattering rate $\Gamma$ versus energy asymmetry $\Delta$, for Coulomb and defect scattering mechanisms, evaluated using Eq. \ref{eq:tau-avg}. Coulomb and defect scatterers show opposite behavior as $\Delta$ changes. Material parameters used for this plot are the same as those in Fig.~\ref{fig:tau}. 
} 
\label{fig:gamma_delta}
\end{figure}

To study the impact of energy asymmetry on the average scattering rate,
we use the results from Fig.~\ref{fig:tau} in Eq. \ref{eq:tau-avg}. The average scattering rate for both Coulomb and defect scattering mechanisms versus $\Delta$ is plotted in Fig.~\ref{fig:gamma_delta}.
We see that as $\Delta$ increases from $0$ to $0.8\,$eV, the average scattering rate increases by 70\% for the case of defect scatterers, whereas it decreases by 30\% for the case of Coulomb scatterers. 
The reason for the unexpected behavior of Coulomb scatterers is the nontrivial behavior of polarizability. Our calculations show that for $\mu=0.3\,$eV, introducing a large energy asymmetry will increase the polarizability of BLG and consequently its dielectric function. This is the effect that outweighs the reduction in Fermi velocity and leads to a smaller average scattering rate overall in a sample dominated by Coulomb scatterers. 

\begin{figure} [ht]
\begin{center}
\includegraphics[width=3.4in]{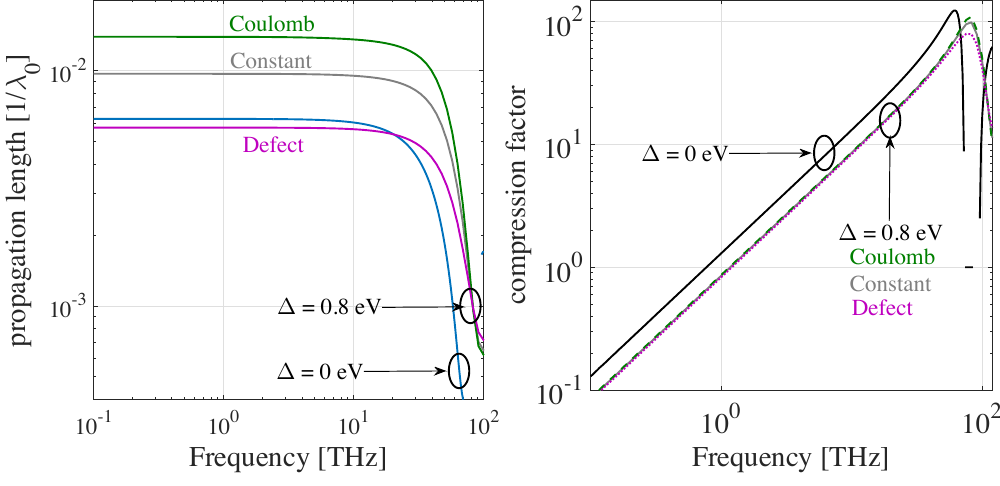}
\end{center}
\caption 
{Propagation length (Left) and compression factor (Right) versus frequency for both zero and finite energy asymmetry i.e. $\Delta=0$ and $\Delta=0.8\,$eV. Three cases of (i) constant scattering rate, (ii) Coulomb-dominated scattering, and (iii) defect-dominated scattering are considered. For the gapless BLG ($\Delta=0$), $\Gamma=20\,$meV is the same for all the three cases.} 
\label{fig:prop}
\end{figure}

Average scattering rate determines the intraband optical absorption and, therefore, the propagation length of plasmons in BLG.
Depending on the dominant scattering mechanism, the propagation length could be improved or degraded.
In some experimental works, defect scattering is believed to be dominant\cite{monteverde2010transport, sarma2010theory}, in some other works, it is Coulomb scattering that limits carrier transport\cite{zhu2009carrier, xiao2010charged}.
For Coulomb-dominated samples, given the specific Fermi level of $0.3\,$eV, introducing a large $\Delta$ will reduce the average scattering rate and consequently increase the plasmon propagation length, while for defect-dominated samples the opposite effect will happen. 
To elucidate the situation, we plot the plasmon propagation length in Fig.~\ref{fig:prop} in the absence and presence of energy asymmetry for
(i) constant scattering rate, (ii) Coulomb-dominated scattering, and (iii) defect-dominated scattering.
For comparison, we choose disorder concentration in a way that for the gapless BLG, all the three cases have the same average scattering rate i.e. $\Gamma=20\,$meV (or equivalently $4.84\times10^{12}\,$1/s). The disorder concentrations are $n_i=9.57\times10^{16}\,$m$^{-2}$ for Coulomb-dominated scattering and $n_d=0.012\times10^{16}\,$m$^{-2}$ for defect-dominated scattering. 
Introducing an energy asymmetry of $\Delta=0.8\,$eV will increase the average scattering rate of defect-dominated case to $\Gamma_d=34\,$meV which will degrade the propagation length relative to the constant scattering case. On the other hands, it will decrease the average scattering rate of Coulomb-dominated case to $\Gamma_d=14\,$meV which will improve the propagation length relative to the constant scattering case.

\begin{acknowledgments}
The authors acknowledge the funding support of National Science Foundation through the grant no. CCF-1565656.
\end{acknowledgments}

\bibliographystyle{apsrev}
\bibliography{report}

\end{document}